\documentclass[aps,pre,showpacs,preprint]{revtex4} 
\usepackage[T1]{fontenc}
\usepackage[english]{babel}
\usepackage{pslatex}

\usepackage{amsmath}
\usepackage{array}
\usepackage{subfigure}
\usepackage{graphicx}

\begin{document}
%\setlength{\tabcolsep}{1.5mm} 
%\setcounter{totalnumber}{4}
%\setcounter{topnumber}{4}
%\setlength{\voffset}{-0cm}
%\setlength{\hoffset}{-0.cm}
%\addtolength{\textheight}{1.1cm}

%%%%%%%%%%%%%%%%%%%%

%% PRIVATE MACROS %%

%%%%%%%%%%%%%%%%%%%%

\newcommand{\figwidth}{1.00\columnwidth}
\newcommand{\eq}[1]{Eq.(\ref{#1})}
\newcommand{\fig}[1]{Fig.~\ref{#1}}
\newcommand{\sect}[1]{Sec.~\ref{#1}}
\newcommand{\avg}[1]{{\langle #1 \rangle}}
\newcommand{\olcite}[1]{Ref.~\onlinecite{#1}}

%%%%%%%%%%%%%%%%%%%%%%%%%%

%% DOCUMENT STARTS HERE %%

%%%%%%%%%%%%%%%%%%%%%%%%%%

\title{The configurational space of colloidal patchy polymers with heterogeneous sequences}

\author{Ivan Coluzza}
\affiliation{Faculty of Physics, University of Vienna, Boltzmanngasse 5, 1090 Vienna, Austria}

\author{Christoph Dellago}
\affiliation{Faculty of Physics, University of Vienna, Boltzmanngasse 5, 1090 Vienna, Austria}

\date{\today}
\begin{abstract}
In this work we characterize the configurational space of a short chain of colloidal particles as function of the range of directional and heterogeneous isotropic interactions. The individual particles forming the chain are colloids decorated with patches that act as interaction sites between them. We show, using computer simulations, that it is possible to sample the relative probability of occurrence of a structure with a sequence in the space of all possible realizations of the chain. The results presented here represent a first attempt to map the space of possible  configurations that a chain of colloidal particles may adopt. Knowledge of such a space is crucial for a possible application of colloidal chains as models for designable self-assembling systems. 
\end{abstract}

\maketitle

\section{Introduction}

Self-assembly is the process by which a substance exclusively driven by non-covalent interactions spontaneously develops into a specific long-lived conformation with a well defined structure~\cite{Lawrence:10.1021/cr00038a018}. 
Self-assembling is common in natural bio-polymers, such as DNA, RNA and, in particular, proteins, that exhibit exceptional self-assembling properties. Proteins, the fundamental building blocks of every living organisms, have the remarkable property that their structure and, therefore, their function is encoded in the one-dimensional sequence of the structural elements that compose them. Although proteins vary strongly in size, structure and function, they are all composed of the same 20 fundamental chemical units called amino acids. 
% IVAN; Here you mentioned that there are 22 amino acids. I have corrected that to 20.
This protein alphabet insures an enormous variety of combinations, however only few sequences will have a well defined stable ground state, each of which consists of complex arrangements of predominantly three types of secondary structures: alpha helices, beta sheets and random coils~\cite{ANFINSEN:1973p2310,Dobson:1998p2346,KeithDunker:2002p2365}. Hence, bio-polymers are a clear reference point for the development of artificial self-assembling systems based on modular subunits. 

Recently, it has been shown~\cite{Coluzza:2011p0020853} that the minimal set of constraints imposed on configurational space by the hydrogen bonds along the protein backbone and the self-avoidance of the residues are sufficient to enable the successful folding of real protein structures. More generally, it is known that  the specific geometry of the backbone and the directionality of the hydrogen bonds ``pre-sculpt'' the configurational space of proteins to the sub-space of currently known proteins structures~\cite{Maritan00,Hoang:2004p1364,Magee06,Banavar:2009p6900}, and allow for the design of sequences capable of folding back to their target structures~\cite{Coluzza:2011p0020853}. We believe that it is possible to extend the concept of configurational space pre-sculpting to generalized chains of particles capable of folding into geometries very different from those observed for natural proteins. However, for the future design of complex structures it is crucial to characterize the variety of  configurations that a given set of interactions can generate. This is equivalent to the still ongoing process of mapping the ``protein universe'', where huge efforts are devoted to finding all protein structures corresponding to all sequences expressed in living organisms~\cite{PDB}. 

In this work, we explore the total phase space of colloidal patchy polymers. Such phase space is the ensemble of all chain configurations defined by a structure and a sequence along the chain of particles. Each particle is decorated by spots, or {\em patches}, with interactions that mimic the directionality of the hydrogen bonds. Such directional interactions confine the configurational space of the synthetic protein analogue to a sub-space of compatible structures. We map the phase space for a range of values of the interaction potential and for two different forms of the interaction potential. We have chosen as subunits patchy particles, because in addition to mimicking the properties of hydrogen bonds, they have a rich ensemble of self-assembling properties~\cite{0953-8984-19-32-323101,Bianchi2011:10.1039/c0cp02296a,PhysRevLett.106.255501}. Experimentally, patchy colloidal particles are spherical particles with a hard core and a radius that varies from the nanometer to the micrometer scale and several methods have been developed for the experimental realization of such particles~\cite{Cho:10.1021/ja0550632,Zhang:2005p143,Mirkin:1996p607}. A beautiful review on the experimental methods to produce patchy particles can be found in the paper of Bianchi et al. ~\cite{Bianchi2011:10.1039/c0cp02296a}.

The results presented in this paper represent a first attempt to link the configurational space of a generic chain of colloidal particles  to simple geometrical parameters of the interaction potential between the particles. Such a space represents a ``phase diagram`` of the designability of the system, because the occurrence probability reflects the propensity of a generic sequence to adopt a given structure. Hence, the larger the space the more versatile is the corresponding model. Below, we will first introduce the details of the model and explain the computational methods used to sample the complex configurational space. Then, we will study the properties of chains of particles with three patches each, where each patch has different interaction ranges relative to the particle-particle interaction, as well as two different functional forms.

\section{Methods}

\subsection{Model}
%--------------------------------------------------------------------------------------
\begin{figure}[ht] 
\begin{center} 
\includegraphics[width=8.5cm]{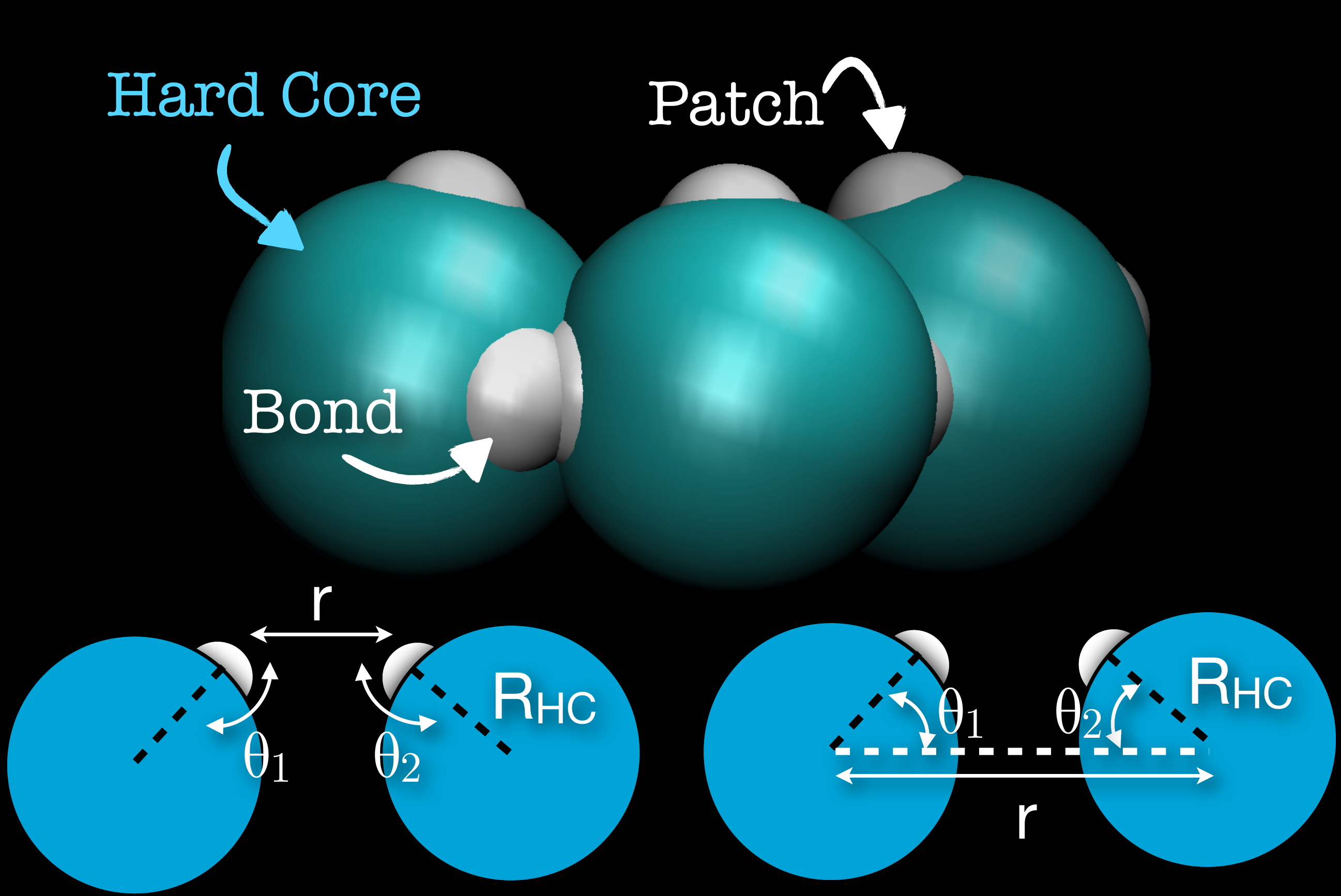}
\end{center} 
\caption{Real-space representation of the backbone of the patchy polymer. The patches are indicated by small white spheres, while the large turquoise spheres represent the hard core of the colloidal particles. If a comparison between this chain and the caterpillar backbone is made~\cite{Coluzza:2011p0020853}, it is easy to see how the patches are reminiscent of the O and H atoms of the protein backbone. 
The insets show a schematic representation of the distance $R$ between the patches and the alignment angles $\theta_1$ and $\theta_2$, which, according to Eq.~\ref{HBond-Energy-1} and Eq.~\ref{HBond-Energy-2} , determine the interactions between patches not involved in the polymer backbone. }
\label{2-patch-wigly} 
\end{figure}
%--------------------------------------------------------------------------------------

The system we study here consists of a single chain of particles, each of which can occur in different types mimicking the various residues of a real protein. Each particle is decorated with 3 patches, two of which are used to link the particle with neighboring particles and form the chain as illustrated in Fig.~\ref{2-patch-wigly}. These connecting patches interact via a simple harmonic spring potential, 
\begin{equation}
E_\text{link}(r)=\frac{\kappa}{2}\,r^2.
\end{equation}
Here, $r$ is the distance between the two anchoring patches and $\kappa=5\:k_{\rm B}T\,R_{HC}^{-2}$, where $R_{HC}$ is the ``hard core'' radius of the colloids. In what follows all the energies are given in units of $k_BT_{\text{ref}}$, where $T_{\text{ref}}$ is a reference temperature that sets the scale of all interactions,  hence in what follows the temperatures do not have units.

The model colloidal polymer studied here is based on the caterpillar model \cite{Coluzza:2011p0020853}. Accordingly, the colloidal particles interact pairwise via a smoothed square well pair potential
\begin{equation}
E_{AB}\left(r\right) =\left\{ \begin{array}{ll} \epsilon_{AB}\left[1-\frac{1}{1 + e^{\left(
r_{\textrm{max}} - r \right)/K}}\right] &\mbox{ if $r>R_{HC},$} \\
\infty &\mbox{ if $r \leq R_{HC},$} \end{array} \right.
\label{Calpha-Energy}
\end{equation}
where the subscripts $A$ and $B$ refer to the type of two interacting particles and $r$ is the distance between their centers. The parameter $K=0.2 R_{HC}$ determines the width of the range over which the potential changes from $\simeq 0$ to $\simeq \epsilon_{AB}$, and $r_{\textrm{max}}$ is the distance at which $E_{AB}(r_{\rm rmax})=\epsilon_{AB}/2$. We have varied $r_{\textrm{max}}$ in the range $r_{\textrm{max}}={2.2,2.5,3.0,4.0,6.0}\;[R_{HC}]$. The parameter $\epsilon_{AB}$, which depends on the types $A$ and $B$ of the interacting particles, determines the depth of the well, i.e., the strength of the interactions. Depending on the particular type of the interacting particles, $\epsilon_{AB}$ can be positive or negative, leading to repulsive or attractive interactions, respectively. Each particle is then assigned a particular identity that will define the interactions with the other particles along the chain. There  is no unique way to choose the values of the $\epsilon_{AB}$, and we will show that with just an alphabet of 2 letters it is possible to produce a wide range of structures. The 2 letters will only distinguish between hydrophilic (P) and hydrophobic particles (H) with $\epsilon_{PP}=0$, $\epsilon_{HH}=-\alpha\,k_BT_{\text{ref}}$, $\epsilon_{HP}=-\alpha/2\,k_BT_{\text{ref}}$. Here $\alpha$ is an adjustable parameter that we will fix to keep the second virial coefficient $B^{AB}_2\sim 1$ constant for the various ranges $r_{\textrm{max}}$ of the potential. The value of $B^{AB}_2$ corresponds to the contribution of $E_{AB}$ in Eq.~\ref{Calpha-Energy} to the total energy and was calculated with numerical integration for each value of  $r_{\textrm{max}}$.

As patch-patch interaction, acting only among the free patches not involved in the links along the chain, we use two sets of directional interactions. The first one is inspired by the hydrogen bond interaction from the  caterpillar model, which is represented by a 10-12 Lennard-Jones type potential with an additional angular dependence that takes into account the directionality of the patch bonds \cite{Irback:2000p1573},
\begin{equation}
E_{P1}=-\epsilon_{P1}\left(\cos{\theta_1}\cos{\theta_2}\right)^\nu
\left[5\left(\frac{\sigma_{P1}}{r}\right)^{12}-
 6\left(\frac{\sigma_{P1}}{r}\right)^{10}\right].
\label{HBond-Energy-1}
\end{equation}
Here, $\nu=2$ as given in ~\cite{Irback:2000p1573}, $r$ is the distance between the patches, and $\theta_1$ and $\theta_2$ are angles between the patches and the centers of their corresponding spheres, as indicated in the left inset of Fig.~\ref{2-patch-wigly}. $\sigma_{P1}$ represents the position of the minimum of the potential; once a value for $\sigma_{P1}$ is fixed then also the range of the potential is determined. The smaller the value of $\sigma_{P1}$ the shorter  the range of the potential will be. For each of the values of $r_{\textrm{max}}$ in Eq.~\ref{Calpha-Energy} we considered short and long range directional interactions summarized below. As for $\alpha$ that defines the  $\epsilon_{AB}$ scale factor in Eq.~\ref{Calpha-Energy}, we determined the  parameter $\epsilon_{P1}$ by imposing that the second virial coefficient $B^{P1}_2$ of the directional potential in Eq.~\ref{HBond-Energy-1} is equal to the second virial coefficient $B^{AB}_2$ for each pair $(r_{\textrm{max}},\sigma_{P1})$. The hard core of the spheres ensures that only the maximum of angular term close to $\pi$ is accessible, i.e. $-\pi$ corresponds to configurations that are sterically inaccessible.

The second type of directional potential is represented by the same square-well like potential defined in Eq.~\ref{Calpha-Energy} modulated by the same angular dependence used for $E_{P1}$ in Eq.~\ref{HBond-Energy-1}, but a different definition of the angles $\theta_1$ and $\theta_2$, 
\begin{equation}
E_{P2}=\left\{ \begin{array}{ll}-\epsilon_{P2}\left(\cos{\theta_1}\cos{\theta_2}\right)^\nu\left[1-\frac{1}{1 + e^{\left(
\sigma_{P2} - r \right)/K}}\right] &\mbox{ if $r>R_{HC}$,} \\
\infty &\mbox{ if $r \leq R_{HC}$,} \end{array} \right.
\label{HBond-Energy-2}
\end{equation}
Here, $\nu=2$, $K=0.2 R_{HC}$, $r$ is now the distance between the centers of the spheres, $\sigma_{P2}$ defines the range of the potential, and $\theta_1$ and $\theta_2$ are now the angles of the projection of each patches  on the vector joining the centers of the interacting spheres, as indicated in the right inset of Fig.~\ref{2-patch-wigly}. We choose a different definition for the distance $r$ in the potential $E_{P2}$, compared to $E_{P1}$, because we compute the same distance for the potential $E_{AB}$ in Eq.~\ref{Calpha-Energy}. As the interaction P2 in Eq.~\ref{HBond-Energy-2} is defined between the centers of the spheres, the range $\sigma_{P2}$ will be longer than $\sigma_{P1}$ from Eq.~\ref{HBond-Energy-1} by exactly 2 colloids radius $R_{HC}$. Again, for each of the values of $r_{\textrm{max}}$ in Eq.~\ref{Calpha-Energy} we considered short and long range directional interactions summarized in Table~\ref{Parameters}. The  parameter $\epsilon_{P2}$ is determined by imposing that the second virial coefficient $B^{P2}_2$ of the directional potential in Eq.~\ref{HBond-Energy-2} is equal to the second virial coefficient $B^{AB}_2$ for each pair $(r_{\textrm{max}},\sigma_{P2})$. The directionality of the patch bonds, encoded in the angular term in Eqs.~\ref{HBond-Energy-1} and ~\ref{HBond-Energy-2} , are each essential to pre-sculpt the conformational space in analogy to the role of hydrogen bonds that characterize the secondary structure elements typical of proteins.

We considered the directional interactions in Eqs.~\ref{HBond-Energy-1} and ~\ref{HBond-Energy-2} as a way to estimate the dependence of configurational space on the different ways of imposing the directionality. Since the directional contribution as well as the window of interaction ranges considered are the same, the only difference between the two interactions is in the shape of the potential. While the interaction in Eq.~\ref{HBond-Energy-1} has a sharp minimum, the square-well profile of the  potential in Eq.~\ref{HBond-Energy-2}, introduces less frustration when combined with the  isotropic interaction of Eq.~\ref{Calpha-Energy}, which has the same functional form. Moreover, there are also experimental reasons to consider the above mentioned interactions. The first potential can represent a short-range directional  attraction on the surface (e.g. Hydrogen Bonds, short DNA), while the second is representative of a class of particles, where the patch is produced by creating a dent in the isotropic interaction.  
\begin{table}[hb!]
\begin{center}
\begin{tabular}{|c||c|c|c|c|c|c|c|c|}
\hline
 Parameter & Set 1 & Set 2 & Set 3 & Set 4 & Set 5 & Set 6 & Set 7 & Set 8 \\
\hline
\hline
 $r_{\textrm{max}}$ $[R_{HC}]$ & 2.2 & 2.5 & 3.0 & 4.0 & 4.0 & 6.0 & 6.0 & 6.0 \\
\hline
$\alpha$ $[k_BT_{\text{ref}}]$ & 2.6 & 1.7 & 1.0 & 0.441 & 0.441 & 0.14 & 0.14 & 0.14 \\
\hline
 $\sigma_{P1}$ $[R_{HC}]$ & 0.2 & 0.2 & 0.2 & 0.2 & 1.0 & 0.2 & 1.0 & 2.0 \\
\hline
$\epsilon_{P1}$ $[k_BT_{\text{ref}}]$ & 5.95 & 5.95 & 5.95 & 5.95 & 4.0 & 5.95 & 4.0 &  3.1 \\
\hline
$\sigma_{P2}$ $[R_{HC}]$ & 2.2 & 2.2 & 2.2 & 2.2 & 3.0 & 2.2 & - & 4.0 \\
\hline
$\epsilon_{P2}$ $[k_BT_{\text{ref}}]$ & 5.0 & 5.0 & 5.0 & 5.0 & 2.7 & 5.0 & - & 1.38 \\
\hline
\end{tabular}

\end{center}
\caption{Values of the potential parameters used in our simulations. It is important to remember that as the range $\sigma_{P2}$ is calculated between the centers of the colloids it will be longer than $\sigma_{P1}$ by exactly 2 colloids radius $R_{HC}$}
\label{Parameters}
\end{table}

\subsection{Simulation methodology} 
\label{sec:design}

The first step in the characterization of the patchy polymer model is the classification of the different structures according to how easy it is to design them. In order to do so, we will compare the {\em designability} of many chain configurations by estimating the number of sequences that fold into each structure. In order to explore this vast space we combine a particle identity mutation  move (see below)  with the standard set of pivot crankshaft and single particle moves used to explore the configurations of chains of particles~\cite{FrenkelSmit}. For relatively short chains it is possible to identify the most recurrent structures and compare their designability using the algorithm described below. We considered first the simple case of a chain of 20 colloidal particles chosen  from an alphabet of size 2 as described in the section II.A.

As we aim to imitate the designability property of natural proteins, we based the identity mutations move on the design scheme successfully used in protein studies~\cite{Coluzza:2011p0020853}. As in the conventional Metropolis scheme, the acceptance of such trial moves depends on the ratio of the Boltzmann weights of the new and old states for each temperature $T$~\cite{FrenkelSmit}. 
However, if this were the only criterion, there would be a tendency to generate homopolymer chains with a low energy, rather than chains that fold selectively into a specific target structure. To ensure a type  composition far from the homopolymer region of sequence space, we sample the equilibrium sequences for the generalized energy function 
\begin{equation}
W=E - \epsilon_{p} \ln N_{P}
\label{equ:generalized}
\end{equation}
where $E$ is the energy of the patchy  polymer defined as the sum of all the contributions calculated according to Eq.~\ref{Calpha-Energy} (as the structure does not change the spring and directional terms are constant), $\epsilon_{p}$ is a scale factor adjusting the relative importance of the two terms in the equation, and $N_{P}$ is the number of permutations that are possible for a given set of particle of given types,
\begin{equation}
N_{P}=\frac{N!}{n_{1}!n_{2}!n_{3}!\cdots n_M!}.
\label{Number of Permutations ALL}
\end{equation}
Here, $N$ is the total number of particles in the chain, $M$ is the number of different particles types, and $n_{1},n_{2}, \cdots, n_M$, etc. are the numbers of particles of type $1, 2, \cdots, M$, respectively. In the present study we have considered an alphabet of only two particles types, namely $H$ and $P$. Hence, it is simply the ratio:
\begin{equation}
N_{P}=\frac{N!}{n_{H}!n_{P}!}.
\label{Number of Permutations}
\end{equation}
While sampling sequence space with the Monte Carlo scheme, we set $\epsilon_{p}$ to a sufficiently high value, $\epsilon_{p} = 5\:k_{\rm B}T$, to generate sequences with a heterogeneous composition. Note that the additional term to the generalized energy of Eq. (\ref{equ:generalized}) is a phenomenological bias that drives the simulation towards the region of heterogeneous sequences.
Although due to this particular bias we may miss some low energy configurations, we believe that an exhaustive sampling of the entire sequence space of each structure is not necessary to compare the designability.

During each simulation we projected the configurational space on the collective variable $Q_S$, which counts the number of contacts between the spheres. We then compute the free energy $F(Q_S)$ as a function of $Q_S$,
\begin{equation}
  F\left(Q_S\right)=-kT\ln\left[P\left(Q_S\right)\right]. \label{equation7}
\end{equation}
Here, $P(Q_S)$ denotes the normalized histogram of the number of sampled conformations with order parameter $Q_S$. In practice, a direct calculation of this histogram is not efficient, since even such short chains tends to be trapped in local minima, especially at low temperatures. To induce escape from these local minima, we made use of the Virtual Move Parallel Tempering Monte Carlo sampling scheme proposed by Coluzza and Frenkel~\cite{Coluzza:2005p596}, based on the Waste Recycling approach~\cite{Frenkel:2004p329}. This scheme is very efficient in sampling both high and low free energy states.

\begin{figure}[ht] 
\begin{center} 
\includegraphics[width=1.0\textwidth]{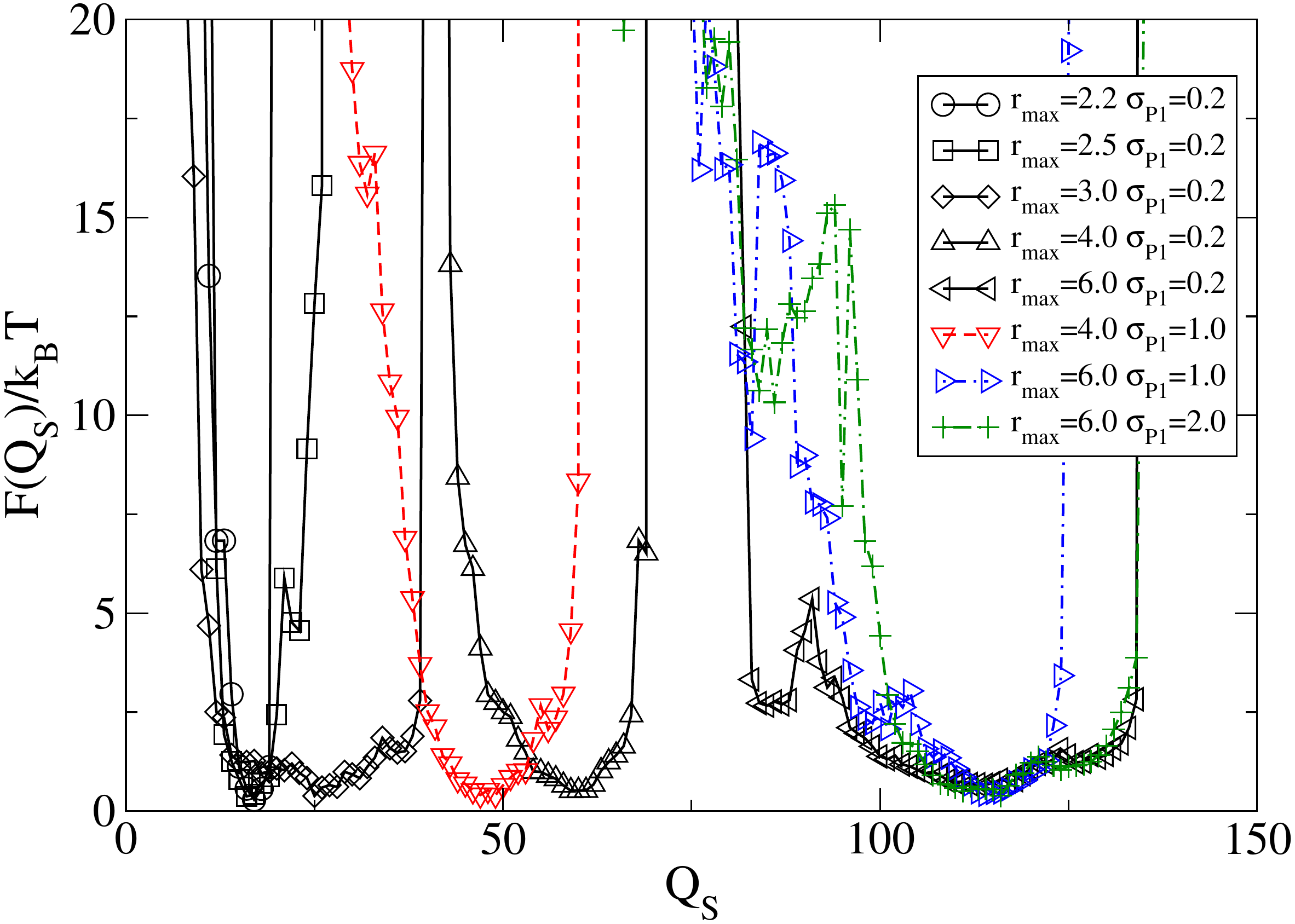}
\end{center} 
\caption{Free energy $F(Q_S)/k_{\rm B}T$  as a function of the number $Q_S$ of contacts among the spheres obtained for a patchy polymer of length $N=20$ with 3 patches per particle interacting with the directional interaction defined in Eq.~\ref{HBond-Energy-1}. The simulations were performed by altering the structure of the chain as well as by mutating the identities of the spheres that intervene in $E_{AB}$ interaction in Eq.~\ref{Calpha-Energy}, at the temperature $T=0.2$. Each curve represents a simulation with a different pair of values $(r_{\textrm{max}},\sigma_{P1})$ as indicated in Table ~\ref{Parameters}. To each pair corresponds a different symbol, however we have grouped the curves with the same range of the directional potential $E_{P1}$ defined in Eq.~\ref{HBond-Energy-1}.}
\label{Fold_Seq_Short2}
\end{figure}

\section{Results}

\begin{figure}[ht] 
\begin{center} 
\includegraphics[width=0.5\textwidth]{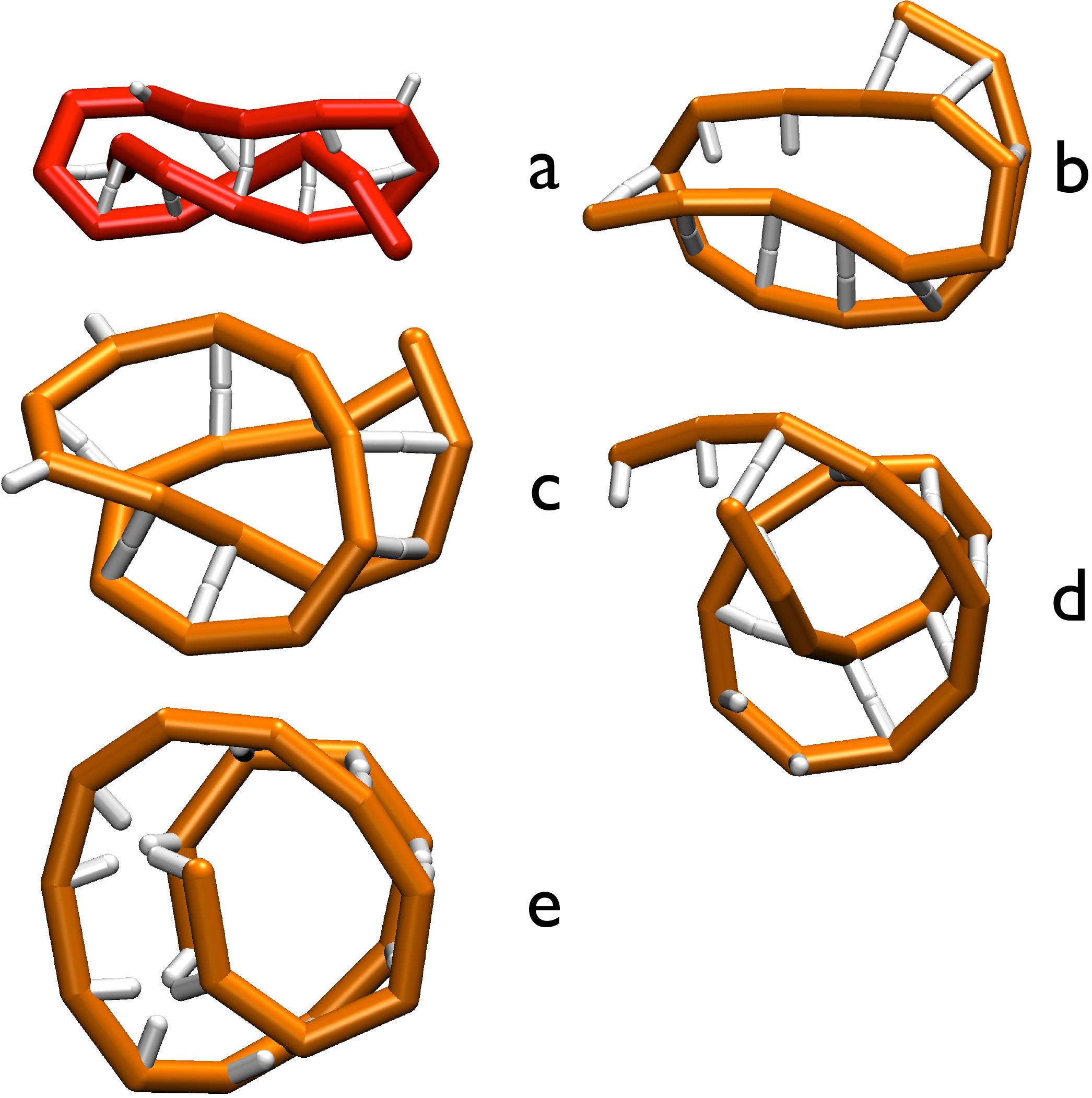}
\end{center} 
\caption{Typical configurations obtained from the simulations of the interaction potential P1 (Eq.~\ref{HBond-Energy-1}). For clarity we have not represented each sphere along the chain, but instead we draw only the backbone as a thick line joining the centers of the spheres. The patches are represented as white stick anchored on the center of the colloid and of length equal to the radius of the colloid, so that the end of the stick corresponds to the position in space of the patch. The structures have been taken from the ensemble of structures corresponding to the minimum of the free energy $F(Q_S)/k_{\rm B}T$ for the combinations of parameters: a) $r_{\text{max}}=2.2\;\sigma_{P1}=0.2$, b) $r_{\text{max}}=3.0\;\sigma_{P1}=0.2$, c) $r_{\text{max}}=4.0\;\sigma_{P1}=0.2$, d) $r_{\text{max}}=6.0\;\sigma_{P1}=0.2$, e) $r_{\text{max}}=6.0\;\sigma_{P1}=1.0$. The configurations have a loop structure that folds on itself very similar to what in proteins is called a beta-sheet. We colored structure (a) in red to highlight the loop-helix nature of the backbone which resembles the helices in Fig.~\ref{Structures_Helices_P1} corresponding to the other set of parameters.}
\label{Structures_Loops_P1}
\end{figure}

\begin{figure}[ht] 
\begin{center} 
\includegraphics[width=0.5\textwidth]{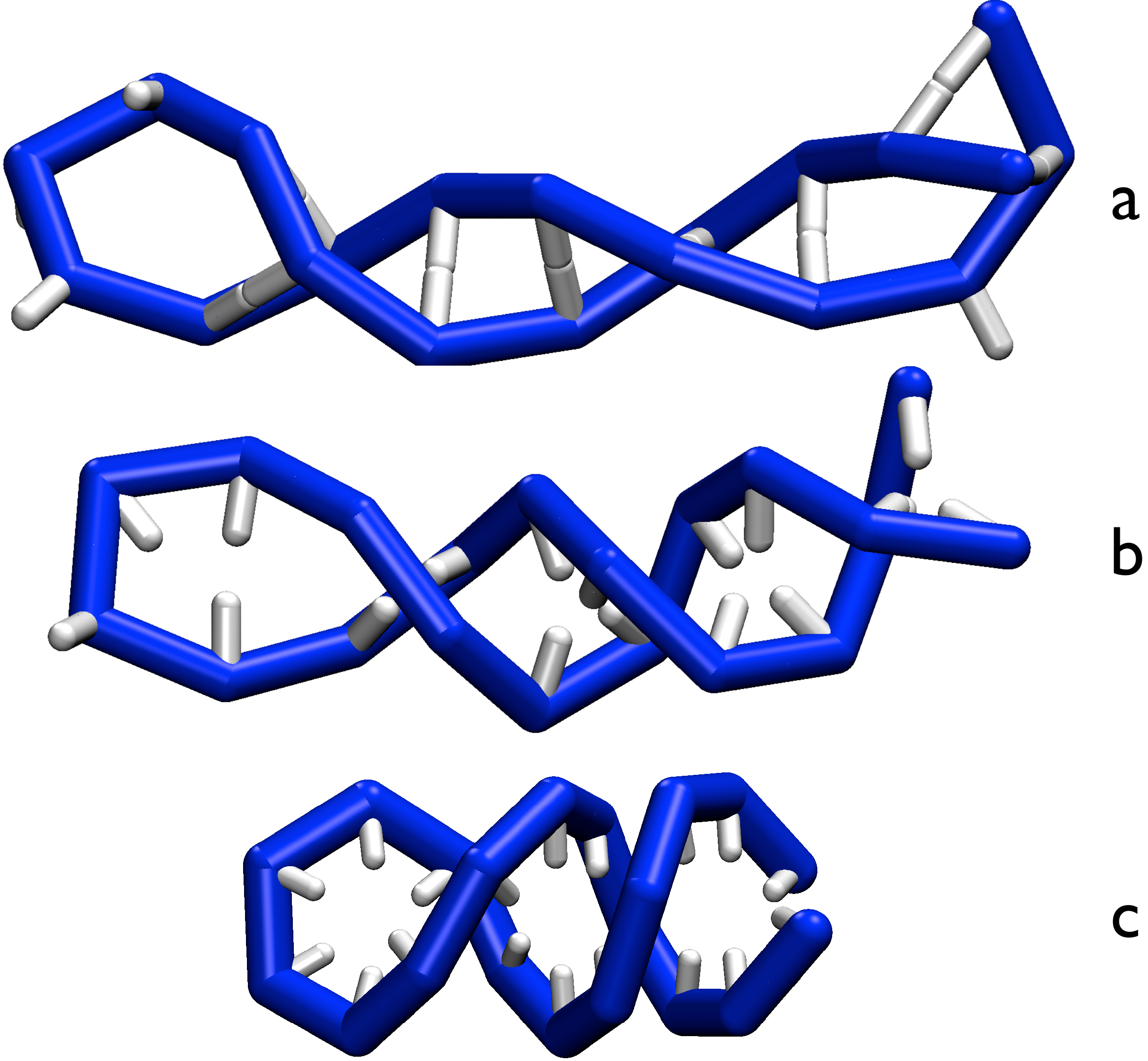}
\end{center} 
\caption{As for the calculations shown in  Fig.~\ref{Structures_Loops_P1} we have isolated the characteristic structures of the minimum free energy $F(Q_S)/k_{\rm B}T$ but now for the combinations of parameters: a) $r_{\text{max}}=2.5\;\sigma_{P1}=0.2$, b) $r_{\text{max}}=4.0\;\sigma_{P1}=1.0$, c) $r_{\text{max}}=6.0\;\sigma_{P1}=2.0$. All the systems exhibit a strong preference for helical configurations of the backbone. To be noted is the shrinking effect that helices show upon increase of the range  $r_{\text{max}}$ of the isotropic interaction.}
\label{Structures_Helices_P1}
\end{figure}

We first carried out a simulation in which we varied both the configuration as well as the sequence of the chain for the patch-patch interaction P1 from Eq.~\ref{HBond-Energy-1}. We considered the combination of parameters in Table ~\ref{Parameters} and we projected the structure space on the collective variable  $Q_S$, which is a count of  the number of contacts between the spheres.  In Fig.~\ref{Fold_Seq_Short2}, we show the free energy profiles $F(Q_S)/k_{\rm B}T$ as a function of $Q_S$ at various combinations of the potentials ranges $r_{\text{max}}$ and $\sigma_{P1}$. From the plot we can see that for   $r_{\text{max}}=2.2, 2.5$ and $\sigma_{P1}=0.2$ the landscapes present a single sharp minimum, indicating that there is a preferred number of contacts for the structures that have high occurrence. As  $r_{\text{max}}$ is increased, the position of the minimum, not surprisingly, shifts  towards higher values of $Q_S$ and it also widens suggesting a richer structural space.
We further characterized the ensemble of configurations by looking at the structures that correspond to the global minimum of each free energy profile. In Figs.~\ref{Structures_Loops_P1} and ~\ref{Structures_Helices_P1} we show a real space representation of such structures. The first aspect that becomes apparent is that there is a wide range of possible structures (it is important to remember that the structures in the minimum are only the most common and not the only possible ones). In addition, we observed the appearance of double helices at the values  $r_{\text{max}}$ and $\sigma_{P1}$  for which the interacting patches sit very close to the half distance  $r_{\text{max}}$ between the centers of the spheres ( the case $r_{\text{max}}=2.2 , \sigma_{P1}=0.2$ is a bit of an exception as it shows only a partial helical configuration). It is important to notice that the pitch of the double helical structures in Fig.~\ref{Structures_Helices_P1} shrinks with increasing values of $r_{\text{max}}$. One could imagine a system where the range of the isotropic interaction is controllable by external parameters (e.g. PH, DNA linkers, depletion), which in turns will allow for the control over the extension of the helix.

\begin{figure}[ht] 
\begin{center} 
\includegraphics[width=1.0\textwidth]{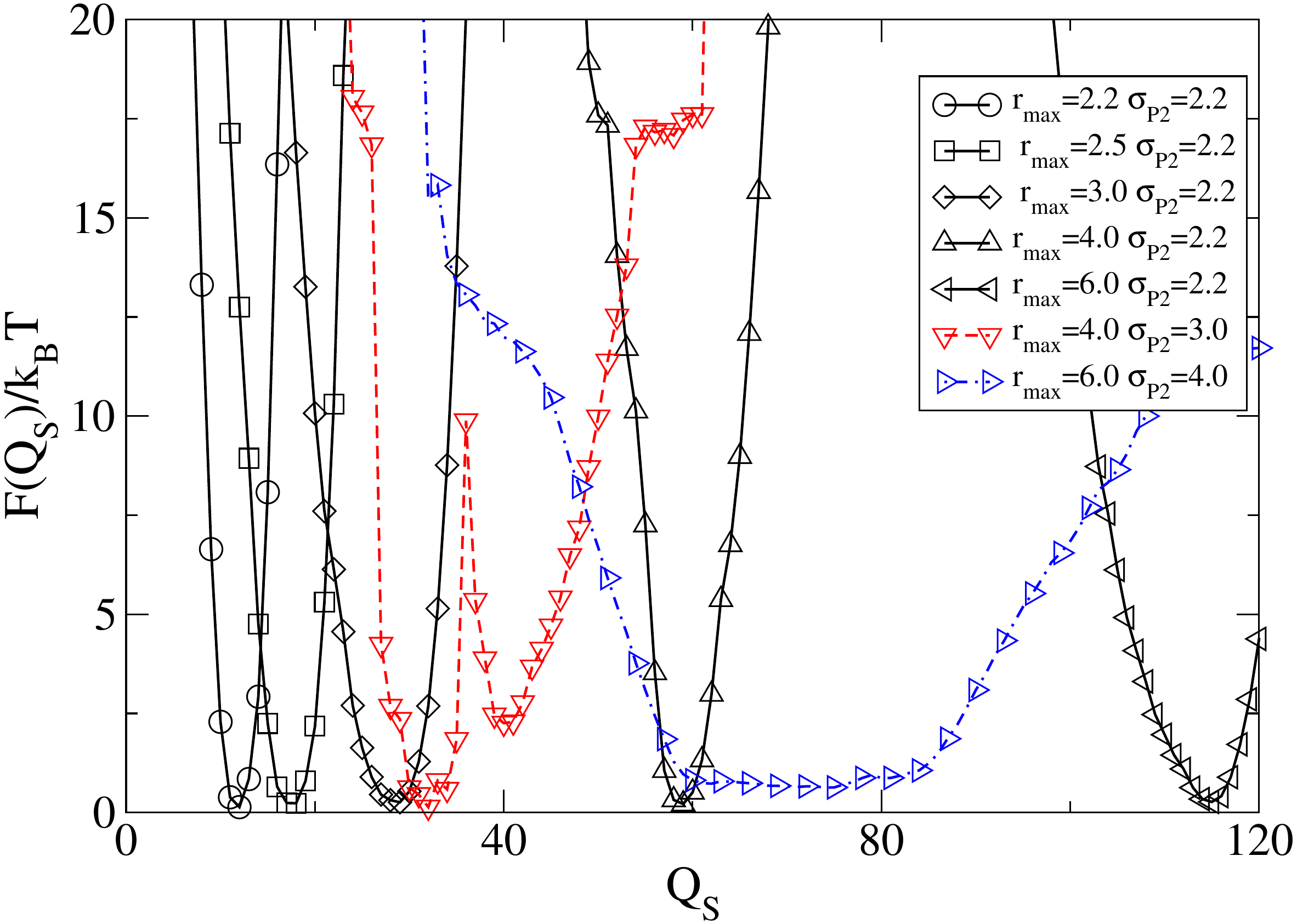}
\end{center} 
\caption{Free energy $F(Q_S)/k_{\rm B}T$  as a function of $Q_S$ for the same simulation conditions and parameters as in Fig.~\ref{Fold_Seq_Short2}, but with the patch-patch interaction from Eq.~\ref{HBond-Energy-2}.  Each curve represents a simulation with a different pair of values $(r_{\textrm{max}},\sigma_{P2})$ as indicated in Table ~\ref{Parameters}. To each pair corresponds a different symbol (see inset), however we have grouped the curves with the same range of the directional potential $E_{P2}$ defined in Eq.~\ref{HBond-Energy-2}. The wide free energy minima observed for the $r_{\textrm{max}}=6.0,\sigma_{P2}=4.0$ case is due to the large fluctuations that the long range potentials can tolerate.}
\label{Fold_Seq_Ptype1}
\end{figure}

\begin{figure}[ht] 
\begin{center} 
\includegraphics[width=0.5\textwidth]{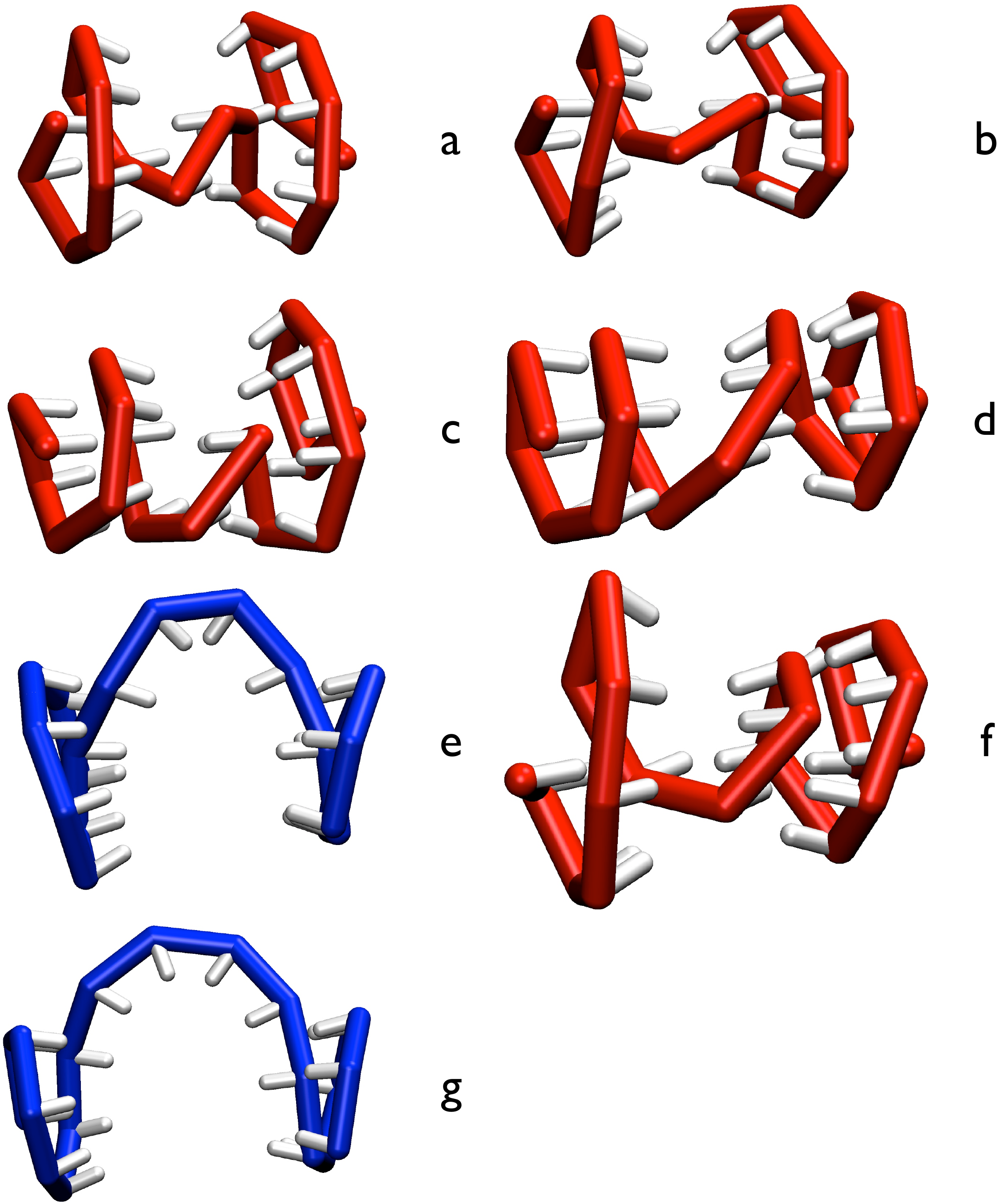}
\end{center} 
\caption{Minimum free energy configurations (see Fig.~\ref{Fold_Seq_Ptype1}), obtained from the simulations of the P2 (Eq.~\ref{HBond-Energy-2}) interaction potential. Each structure is representative of different combinations of the potentials range: a) $r_{\text{max}}=2.2\;\sigma_{P2}=2.2$, b) $r_{\text{max}}=2.5\;\sigma_{P2}=2.2$, c) $r_{\text{max}}=3.0\;\sigma_{P2}=2.2$, d) $r_{\text{max}}=4.0\;\sigma_{P2}=2.2$, e) $r_{\text{max}}=4.0\;\sigma_{P2}=2.0$, f) $r_{\text{max}}=6.0\;\sigma_{P2}=2.2$, g) $r_{\text{max}}=6.0\;\sigma_{P2}=3.0$. We colored structures (e) and (g) in blue because they correspond to the same set of parameters that for the P1 interaction potential gave rise to the structures in Fig.~\ref{Structures_Helices_P1}. Unexpectedly all the structures are almost identical (with an average root mean square distance between the centers of the spheres of 1 $R_{HC}$). The common backbone arrangement appears to be a distorted helix where the patches interact via the tails of the potential in Eq.~\ref{HBond-Energy-2}.  }
\label{Structures_Loops_P2}
\end{figure}

We now consider the second interaction potential P2 from Eq.~\ref{HBond-Energy-2}. We repeated the same procedure described above for the P1 interaction potential. In Fig.~\ref{Fold_Seq_Ptype1}, we plot the free energies $F(Q_S)/k_{\rm B}T$ for the values of the parameters $r_{\textrm{max}},\sigma_{P2}$ in Table ~\ref{Parameters}, as a function of the same collective variable $Q_S$. As before we observe a clear trend that the value of $Q_S$ at the minimum increases for increasing values of $r_{\textrm{max}}$. However, if we look at the actual structures we noticed that they are all very close to each other (see Fig.~\ref{Structures_Loops_P2}) indicating that the displacement of the minimum is just an effect of counting more neighbors per particle. This is true also for the longer ranges of the directional potential, that show for $\sigma_{P2}=3.0$ a double minimum and for  $\sigma_{P2}=4.0$ a very wide basin. The latter is the result of large fluctuations that the long range potentials can accommodate. It is striking to obtain such a dramatic preference for a specific configuration just by slightly altering the form of the interaction potential. By using the same form of the potentials for both isotropic and directional contributions we have reduced the frustration between the two potentials, allowing the system to find an optimal structure for all sets of parameters. We stress that we have chosen the values of the parameters in order to fix the second viral coefficient of all the terms of the potential constant across all simulations. Slightly different configurations occur when the range of the directional interaction $\sigma_{P2}$ is half of range $r_{\textrm{max}}$ the isotropic  interaction (See (e) and (g) of Fig.~\ref{Structures_Loops_P2}). The main difference between the configurations (e) and (g) and the others is in the central loop of the backbone, that for larger values of $\sigma_{P2}$ expands into an extended arch. In particular the arch allows for large compressions of the molecule that results in the wide minima for $\sigma_{P2}=3.0$ in Fig.~\ref{Fold_Seq_Ptype1}.

% \begin{figure}[ht] 
% \begin{center} 
% \includegraphics[width=1.0\textwidth]{Loops_B_P2.pdf}
% \end{center} 
% \caption{Configurations that do not follow the trend of what observed in most cases for the  P2 (Eq.~\ref{HBond-Energy-1}) interaction potential. The structures have been taken from the ensemble of structures corresponding to the minimum of the free energy $F(Q_S)/k_{\rm B}T$ for the combinations of parameters: a) $r_{\text{max}}=4.0\;\sigma_{P2}=2.0$, b) $r_{\text{max}}=6.0\;\sigma_{P2}=3.0$.}
% \label{Structures_Loops_B_P2}
% \end{figure}

\section{Summary and Discussion}

In this work, we present a methodology to explore the configurational space of a chain of colloidal particles decorated with a single patch. The particles  interact through an isotropic potential that depends on the identity of the particles as well as through a direction dependent patch-patch interaction. We developed a methodology to sample the probability of observing a structure with many different sequences therefore allowing for a comparison of the relative probability of observing each configuration. The complexity of such a space gives an idea of how a particular model is suitable for the design of self-assembling structures. We then defined two different forms for the interaction potential and for each of the potentials we considered a wide range of values for the parameters of the potential. From the analysis of the projection of such a space on two collective variables, we identified a wide range of structures with high probability of being observed. 

Our analysis demonstrates that the combination of interactions, with which the chain of colloids is decorated, is capable of shaping the configurational space to a reduced ensemble of configurations. Furthermore, we have shown that the reduced ensemble can be dramatically altered by changing few parameters of the interaction potential (e.g., the interaction range). Such control over the size and the structure heterogeneity of the configurational space makes the chain of patchy-colloids an ideal candidate for the experimental realization of new self-assembling systems. Moreover, we have provided initial but strong indications that, by inducing frustration between the directional (Eq.~\ref{HBond-Energy-1}) and the isotropic potential  the chain is forced to adopt a wider range of structures, compared to the scenario where both potentials shared the same functional form (Eq.~\ref{HBond-Energy-2}). In particular, for specific parameters the backbone of the chain has a strong preference for ordered double helices with a pitch controllable by the range of the isotropic interaction. 

Our results provide an important starting point for the realization of designable chains of patchy colloids. The methodology used here can be easily extended to an arbitrary set of interactions and provide, as a result, a set of target structure for the design of chains of patchy colloids.

\begin{acknowledgments}
IC would like to thank Prof. Erik Reimhult, Dr. Peter van Oostrum and Dr. Ronald Zirbs for the enlightening scientific discussions. We acknowledge support from the Austrian Science Fund (FWF) within the SFB ViCoM (F 41). All simulations presented in this paper were carried out on the Vienna Scientific Cluster (VSC).
\end{acknowledgments}

\bibliography{Biblio-cp14}
\clearpage

\end{document}